\documentclass{article}
\usepackage{amssymb}
\begin{document}
\title{Transformation of a wavefunction under changes of reference frame}

\author{G.F.\ Torres del Castillo \\ Departamento de F\'isica Matem\'atica, Instituto de Ciencias \\
Universidad Aut\'onoma de Puebla, 72570 Puebla, Pue., M\'exico \\[2ex]
B.C.\ N\'ajera Salazar \\ Facultad de Ciencias F\'isico Matem\'aticas \\ Universidad Aut\'onoma de Puebla, 72570 Puebla, Pue., M\'exico}

\maketitle

\begin{abstract}
A simple procedure to derive the transformation of a wavefunction under a change of reference frame is applied to some examples and its relation with the transformation of the Hamilton principal function is studied.
\end{abstract}

\noindent PACS numbers: 03.65.Ca; 45.20.Jj
\section{Introduction}
Usually, the wavefunction, employed in the non-relativistic quantum mechanics, and the Hamilton principal function, appearing in the Hamilton--Jacobi equation of classical mechanics, behave as scalar fields. For instance, if one starts with the Schr\"odinger equation written in Cartesian coordinates, its expression in any other coordinate system is obtained by replacing the partial derivatives of the wavefunction with respect to the Cartesian coordinates by its derivatives with respect to the new coordinates. However, in the case of certain transformations, such as the Galilean transformations, the wavefunction acquires an extra phase factor and, similarly, the Hamilton principal function requires an additional term.

One way of finding the transformation law for a wavefunction under a change of reference frame, applicable to the cases where the transformations of interest form a continuous group, consists in finding first the infinitesimal generators of the action of the group on the wavefunctions; then, with the aid of the exponential map, the elements of the group can be constructed and, making use of the BCH formula, the desired transformations can be expressed in a convenient manner (see, e.g., Refs.\ \cite{GY,Pa,JE,Do}).

Another approach consists in assuming that the Hamiltonian transforms into some specific operator under the change of frame being considered, and then looking for a transformation of the wavefunction such that a solution of the Schrödinger equation in the initial frame is mapped into a solution of the Schrödinger equation in the second frame (see, e.g., Refs.\ \cite{EM,vG}). In this approach, it is not necessary to consider a continuous group of transformations, but one has to postulate the form of the new Hamiltonian.

In this paper we apply a simple method to find the operator that represents the effect of a change of frame on the state vectors (or on the wavefunctions), without having to impose from the start some specific form for the transformed Hamiltonian. Furthermore, this method is applicable to transformations that do not belong to a continuous group, and we do not have to deal with ``infinitesimal'' transformations.

In Section 2 we show how one can readily obtain the operator that represents a change of frame on the state vectors, presenting several examples. In Section 3 we show that the phase factors appearing in the transformations of the wavefunctions, obtained in Section 2, are given by ${\rm e}^{- {\rm i} F_{1}/\hbar}$, where $- F_{1}$ is the term that has to be added to the Hamilton principal function in the change of frame under consideration.

\section{Transformation of the wavefunctions}
In the context of the non-relativistic quantum mechanics we consider a transformation given by a {\em unitary}\/ operator, $U$, defined by the conditions
\begin{equation}
U x_{i} U^{-1} = X_{i}(x_{j}, t), \qquad U p_{i} U^{-1} = P_{i}(p_{j}, t), \label{3.1}
\end{equation}
where the $x_{i}$ and $p_{i}$ are Hermitian operators representing the Cartesian coordinates and momenta, the $X_{i}$ are given functions of $x_{j}$ and $t$, and the $P_{i}$ are given functions of $p_{j}$ and $t$. For example, for a Galilean transformation
\begin{equation}
{\bf X} = {\bf x} - {\bf V} t, \qquad  {\bf P} = {\bf p} - m {\bf V}, \label{gal}
\end{equation}
where $m$ is the mass of the particle being considered, and ${\bf V}$ is a constant vector, corresponding to the velocity of the boost. (In order to facilitate the comparison with the results of previous works, we consider {\em active}\/ transformations.)

It should be noticed that Eqs.\ (\ref{3.1}) define $U$ up to a phase factor that depends on $t$ only (see the examples below).

The state of the system is transformed according to
\begin{equation}
| \psi' \rangle = U | \psi \rangle \label{3.2}
\end{equation}
and a straightforward computation shows that $U$ maps any solution of the Schr\"odinger equation
\[
{\rm i} \hbar \frac{{\rm d} | \psi \rangle}{{\rm d} t} = H | \psi \rangle
\]
into a solution of
\[
{\rm i} \hbar \frac{{\rm d} | \psi' \rangle}{{\rm d} t} = K | \psi' \rangle
\]
if
\begin{equation}
K = U H U^{-1} + {\rm i} \hbar \frac{{\rm d} U}{{\rm d} t} U^{-1}. \label{3.3}
\end{equation}
This last equation shows that if $U$ depends explicitly on the time, then the Hamiltonian does not transform following the simple rule $H \mapsto U H U^{-1}$ (cf.\ Ref.\ \cite{GYb}). (Note that we are working in the Schr\"odinger picture.) If the arbitrary phase factor contained in $U$ can be chosen in such a way that $K = H$, then we say that $H$ is invariant under $U$ \cite{JE}.

Let $| {\bf x}_{0} \rangle$ and $| {\bf p}_{0} \rangle$ be eigenstates of the position and momentum operators ${\bf x}$ and ${\bf p}$, respectively (with ${\bf x} | {\bf x}_{0} \rangle = {\bf x}_{0} | {\bf x}_{0} \rangle$ and ${\bf p} | {\bf p}_{0} \rangle = {\bf p}_{0} | {\bf p}_{0} \rangle$) then, making use of Eqs.\ (\ref{3.1}) we have
\[
{\bf x} U^{-1} | {\bf x}_{0} \rangle = U^{-1} {\bf X}({\bf x}, t) | {\bf x}_{0} \rangle = {\bf X}({\bf x}_{0}, t) \, U^{-1} | {\bf x}_{0} \rangle,
\]
which means that $U^{-1} | {\bf x}_{0} \rangle$ is an eigenstate of ${\bf x}$ with eigenvalue ${\bf X}({\bf x}_{0}, t)$, thus
\begin{equation}
U^{-1} | {\bf x}_{0} \rangle = {\rm e}^{{\rm i} \alpha/\hbar} | {\bf X}({\bf x}_{0}, t) \rangle, \label{3.1.2}
\end{equation}
where $\alpha$ is some real number, which may depend on ${\bf x}_{0}$, $t$, and the parameters contained in $U$. This last equation, together with (\ref{3.2}), imply that a wavefunction transforms according to
\begin{equation}
\psi'({\bf x}_{0}) = \langle {\bf x}_{0} | U | \psi \rangle = {\rm e}^{- {\rm i} \alpha/\hbar} \langle {\bf X}({\bf x}_{0}, t) | \psi \rangle = {\rm e}^{- {\rm i} \alpha/\hbar} \psi\big( {\bf X}({\bf x}_{0}, t) \big). \label{wftr}
\end{equation}
As we shall see in the examples below, in some cases $\alpha$ is different from zero.

In a similar manner, from Eqs.\ (\ref{3.1}) it follows that
\begin{equation}
U^{-1} | {\bf p}_{0} \rangle = {\rm e}^{{\rm i} \beta/\hbar} | {\bf P}({\bf p}_{0}, t) \rangle, \label{3.1.3}
\end{equation}
where $\beta$ is some real number, which may depend on ${\bf p}_{0}$, $t$, and the parameters contained in $U$. In order to determine the values of $\alpha$ and $\beta$ we form the scalar product $\langle {\bf x}_{0} | U U^{-1} | {\bf p}_{0} \rangle = \langle {\bf x}_{0} | {\bf p}_{0} \rangle = (2\pi \hbar)^{-3/2} \exp ({\rm i} {\bf p}_{0} \cdot {\bf x}_{0}/\hbar)$, which, by virtue of (\ref{3.1.2}), (\ref{3.1.3}) and the unitarity of $U$, must coincide with
\[
(2\pi \hbar)^{-3/2} {\rm e}^{{\rm i} (\beta - \alpha)/\hbar} \langle {\bf X}({\bf x}_{0}, t) | {\bf P}({\bf p}_{0}, t) \rangle = (2\pi \hbar)^{-3/2} \exp \frac{{\rm i}}{\hbar} \big[ \beta - \alpha + {\bf P}({\bf p}_{0}, t) \cdot {\bf X}({\bf x}_{0}, t) \big].
\]
Hence,
\begin{equation}
{\bf p}_{0} \cdot {\bf x}_{0} = \beta - \alpha + {\bf P}({\bf p}_{0}, t) \cdot {\bf X}({\bf x}_{0}, t). \label{bas}
\end{equation}
In the following subsections we consider several applications of the basic formula (\ref{bas}).

\subsection{Spatial translations}
A relatively simple and common example of a change of frame corresponds to translations. We include it because it serves to illustrate the method and because some results will be employed below.

A spatial translation by a constant vector ${\bf a}$ can be defined by Eqs.\ (\ref{3.1}) with
\begin{equation}
{\bf X} = {\bf x} - {\bf a}, \qquad {\bf P} = {\bf p}. \label{tra}
\end{equation}
Then, from Eq.\ (\ref{bas}) we obtain ${\bf p}_{0} \cdot {\bf x}_{0} = \beta - \alpha + {\bf p}_{0} \cdot ({\bf x}_{0} - {\bf a})$, i.e.,
\[
\alpha = \beta - {\bf p}_{0} \cdot {\bf a}.
\]
Hence, taking into account that $\alpha$ may depend on ${\bf x}_{0}$, and $\beta$ may depend on ${\bf p}_{0}$, we conclude that
\begin{equation}
\alpha = \chi(t), \qquad \beta = {\bf p}_{0} \cdot {\bf a} + \chi(t), \label{ab2.1}
\end{equation}
where $\chi(t)$ is a real-valued function of $t$ only. Substituting the expression for $\beta$ into Eq.\ (\ref{3.1.3}), making use of (\ref{tra}), we obtain
\[
U^{-1} | {\bf p}_{0} \rangle = {\rm e}^{{\rm i} \chi/\hbar} \, {\rm e}^{{\rm i}  {\bf p}_{0} \cdot {\bf a}/\hbar} | {\bf p}_{0} \rangle = {\rm e}^{{\rm i} \chi/\hbar} \, {\rm e}^{{\rm i}  {\bf p} \cdot {\bf a}/\hbar} | {\bf p}_{0} \rangle,
\]
which amounts to
\begin{equation}
U^{-1} = {\rm e}^{{\rm i} \chi/\hbar} \, {\rm e}^{{\rm i}  {\bf p} \cdot {\bf a}/\hbar}. \label{traop}
\end{equation}
Substituting (\ref{traop}) and the first equation (\ref{tra}) into (\ref{3.1.2}) we obtain the well-known relation
\begin{equation}
{\rm e}^{{\rm i}  {\bf p} \cdot {\bf a}/\hbar} | {\bf x}_{0} \rangle = | {\bf x}_{0} - {\bf a} \rangle \label{trax}
\end{equation}
(note that the phase factor ${\rm e}^{{\rm i} \chi/\hbar}$ cancels out).

On the other hand, from Eqs.\ (\ref{3.3}) and (\ref{traop}) we find that, in the present case,
\begin{equation}
K = U H U^{-1} + \frac{{\rm d} \chi}{{\rm d} t}, \label{h2.1}
\end{equation}
so that, if, for example,
\begin{equation}
H = \frac{{\bf p}^{2}}{2m} - {\bf F} \cdot {\bf x}, \label{unif}
\end{equation}
where ${\bf F}$ is a constant vector, corresponding to a particle subject to a constant force ${\bf F}$, then [see (\ref{3.1}) and (\ref{tra})]
\[
K = U \left( \frac{{\bf p}^{2}}{2m} - {\bf F} \cdot {\bf x} \right) U^{-1} + \frac{{\rm d} \chi}{{\rm d} t} = \frac{{\bf p}^{2}}{2m} - {\bf F} \cdot ({\bf x} - {\bf a}) + \frac{{\rm d} \chi}{{\rm d} t} = H + {\bf F} \cdot {\bf a} + \frac{{\rm d} \chi}{{\rm d} t}.
\]
{\em If}\/ we demand that $K = H$ (which is reasonable, since the particle is in a uniform field of force, and $U$ represents a translation), we have to choose $\chi = - {\bf F} \cdot {\bf a} t$ and, according to (\ref{wftr}), the wavefunctions must transform as
\begin{equation}
\psi'({\bf x}_{0}) = {\rm e}^{{\rm i} {\bf F} \cdot {\bf a} t/\hbar} \psi({\bf x}_{0} - {\bf a}). \label{2.1.2}
\end{equation}
(Note that with this choice for $\chi$, according to Eq.\ (\ref{traop}), the operator corresponding to translations is $U = {\rm e}^{- {\rm i}  ({\bf p} - {\bf F} t) \cdot {\bf a}/\hbar}$, which involves the conserved operator ${\bf p} - {\bf F} t$ \cite{JE}.)

\subsection{Translations in the momentum}
Even though it is not a change of frame, we shall consider a ``translation'' in the momentum, defined by ${\bf X} = {\bf x}$, ${\bf P} = {\bf p} - {\bf b}$, where ${\bf b}$ is a constant vector. In this case Eq.\ (\ref{bas}) gives ${\bf p}_{0} \cdot {\bf x}_{0} = \beta - \alpha + ({\bf p}_{0} - {\bf b}) \cdot {\bf x}_{0}$, which leads to
\begin{equation}
\alpha = - {\bf b} \cdot {\bf x}_{0} + \chi(t), \qquad \beta = \chi(t), \label{ab2.2}
\end{equation}
where $\chi(t)$ is a real-valued function of $t$ only. Then, from Eq.\ (\ref{3.1.2}) we obtain
\[
U^{-1} | {\bf x}_{0} \rangle = {\rm e}^{{\rm i} \chi/\hbar} {\rm e}^{- {\rm i} {\bf b} \cdot {\bf x}_{0}/\hbar} | {\bf x}_{0} \rangle = {\rm e}^{{\rm i} \chi/\hbar} {\rm e}^{- {\rm i} {\bf b} \cdot {\bf x}/\hbar} | {\bf x}_{0} \rangle
\]
which means that
\begin{equation}
U^{-1} = {\rm e}^{{\rm i} \chi/\hbar} {\rm e}^{- {\rm i} {\bf b} \cdot {\bf x}/\hbar} \label{trm}
\end{equation}
and from Eq.\ (\ref{3.1.3}) we have $U^{-1} | {\bf p}_{0} \rangle = {\rm e}^{{\rm i} \chi/\hbar} | {\bf p}_{0} - {\bf b} \rangle$, i.e.,
\begin{equation}
{\rm e}^{- {\rm i}  {\bf b} \cdot {\bf x}/\hbar} | {\bf p}_{0} \rangle = | {\bf p}_{0} - {\bf b} \rangle. \label{trap}
\end{equation}

Another useful formula follows from the second equation in (\ref{3.1}): $U {\bf p} U^{-1} = {\bf p} - {\bf b}$ or, equivalently,
\begin{equation}
{\rm e}^{{\rm i} {\bf b} \cdot {\bf x}/\hbar} {\bf p} {\rm e}^{- {\rm i} {\bf b} \cdot {\bf x}/\hbar} = {\bf p} - {\bf b}. \label{2.2.2}
\end{equation}
It may be noticed that Eqs.\ (\ref{trap}) and (\ref{2.2.2}) do not contain the function $\chi$.

\subsection{Galilean transformations}
In the case of the Galilean transformations the functions ${\bf X}({\bf x}, t)$ and ${\bf P}({\bf p}, t)$ are given by ${\bf X}({\bf x}, t) = {\bf x} - {\bf V} t$, ${\bf P}({\bf p}, t) = {\bf p} - m {\bf V}$ [see Eqs.\ (\ref{gal})]. Then, Eq.\ (\ref{bas}) becomes
\[
{\bf p}_{0} \cdot {\bf x}_{0} = \beta - \alpha + ({\bf p}_{0} - m {\bf V}) \cdot ({\bf x}_{0} - {\bf V} t),
\]
that is
\[
\alpha + m {\bf V} \cdot {\bf x}_{0} - {\textstyle \frac{1}{2}} m V^{2} t = \beta - {\bf p}_{0} \cdot {\bf V} t + {\textstyle \frac{1}{2}} m V^{2} t,
\]
which implies that
\begin{equation}
\alpha =  - m {\bf V} \cdot {\bf x}_{0} + {\textstyle \frac{1}{2}} m V^{2} t + \chi(t), \qquad \beta = {\bf p}_{0} \cdot {\bf V} t - {\textstyle \frac{1}{2}} m V^{2} t + \chi(t), \label{ab2.3}
\end{equation}
where $\chi(t)$ is some real-valued function of $t$ only. Hence, according to (\ref{3.1.2}), we have
\begin{equation}
U^{-1} | {\bf x}_{0} \rangle = \exp ({\rm i}/\hbar) \big[ - m {\bf V} \cdot {\bf x}_{0} + {\textstyle \frac{1}{2}} m V^{2} t + \chi(t) \big] \, | {\bf x}_{0} - {\bf V} t \rangle, \label{2.2.1}
\end{equation}
which can also be expressed as [see Eq.\ (\ref{trax})]
\[
\exp ({\rm i}/\hbar) \big[ - m {\bf V} \cdot {\bf x}_{0} + {\textstyle \frac{1}{2}} m V^{2} t + \chi(t) \big] {\rm e}^{{\rm i} {\bf p} \cdot {\bf V} t/\hbar} \, | {\bf x}_{0} \rangle
\]
or, equivalently,
\[
\exp ({\rm i}/\hbar) \big[ {\textstyle \frac{1}{2}} m V^{2} t + \chi(t) \big] \, {\rm e}^{{\rm i} {\bf p} \cdot {\bf V} t/\hbar} \, {\rm e}^{- {\rm i} m {\bf V} \cdot {\bf x}/\hbar} \, | {\bf x}_{0} \rangle
\]
and, therefore,
\begin{equation}
U^{-1} = \exp ({\rm i}/\hbar) \big[ {\textstyle \frac{1}{2}} m V^{2} t + \chi(t) \big] \, {\rm e}^{{\rm i} {\bf p} \cdot {\bf V} t/\hbar} \, {\rm e}^{- {\rm i} m {\bf V} \cdot {\bf x}/\hbar}. \label{2.2.5}
\end{equation}

As in Section 2.1, we can determine the function $\chi$ if we impose some specific relation between the Hamiltonians $H$ and $K$ [see Eq.\ (\ref{3.3})]. Substituting (\ref{2.2.5}) into (\ref{3.3}), with the aid of (\ref{2.2.2}), we find
\begin{equation}
K = U H U^{-1} - \frac{1}{2} m V^{2} + {\bf p} \cdot {\bf V} + \frac{{\rm d} \chi}{{\rm d} t} \label{h2.3}
\end{equation}
(cf.\ Ref.\ \cite{SW}). Thus, if we take $H = {\bf p}^{2}/2m$, then
\[
K = \frac{1}{2m} ({\bf p} - m {\bf V})^{2}  - \frac{1}{2} m V^{2} + {\bf p} \cdot {\bf V} + \frac{{\rm d} \chi}{{\rm d} t},
\]
which coincides with $H$ if $\chi = 0$. Other Hamiltonians are also invariant under the Galilean transformations, with the appropriate choice of $\chi$ \cite{JE}.

If one does not allow for the presence of a phase factor ${\rm e}^{{\rm i} \chi/\hbar}$ in $U^{-1}$ one arrives at the wrong conclusion that only the Hamiltonian of a free particle is invariant under the Galilean transformations \cite{JP}.

\subsection{Constant acceleration}
Now we consider the effect of a constant acceleration, ${\bf a}$, which corresponds to
\[
{\bf X} = {\bf x} - {\textstyle \frac{1}{2}} {\bf a} t^{2}, \qquad {\bf P} = {\bf p} - m {\bf a} t,
\]
Substituting these expressions into Eq.\ (\ref{bas}) we have ${\bf p}_{0} \cdot {\bf x}_{0} = \beta - \alpha + ({\bf p}_{0} - m {\bf a} t) \cdot ({\bf x}_{0} - {\textstyle \frac{1}{2}} {\bf a} t^{2})$, which implies that
\begin{equation}
\alpha = - m {\bf a} t \cdot {\bf x}_{0} + {\textstyle \frac{1}{6}} m a^{2} t^{3} + \chi(t), \qquad \beta = {\textstyle \frac{1}{2}} {\bf p}_{0} \cdot {\bf a} t^{2} - {\textstyle \frac{1}{3}} m a^{2} t^{3} + \chi(t), \label{ab2.4}
\end{equation}
where $\chi(t)$ is a function of $t$ only, and we have included the term ${\textstyle \frac{1}{6}} m a^{2} t^{3}$ into $\alpha$ for later convenience.

The expression of the operator $U^{-1}$ can be obtained by calculating $U^{-1} | {\bf x}_{0} \rangle$, following the same steps as in Section 2.3. Alternatively, we can start by considering the action of $U^{-1}$ on $| {\bf p}_{0} \rangle$. From Eqs.\ (\ref{3.1.3}), (\ref{ab2.4}), and (\ref{trap}) we find that
\begin{eqnarray*}
U^{-1} | {\bf p}_{0} \rangle & = & \exp ({\rm i}/\hbar) \left[ {\textstyle \frac{1}{2}} {\bf p}_{0} \cdot {\bf a} t^{2} - {\textstyle \frac{1}{3}} m a^{2} t^{3} + \chi(t) \right] |{\bf p}_{0} - m {\bf a} t \rangle  \\
& = & \exp ({\rm i}/\hbar) \left[ {\textstyle \frac{1}{2}} ({\bf p} + m {\bf a} t) \cdot {\bf a} t^{2} - {\textstyle \frac{1}{3}} m a^{2} t^{3} + \chi(t) \right] |{\bf p}_{0} - m {\bf a} t \rangle  \\
& = & \exp ({\rm i}/\hbar) \left[ {\textstyle \frac{1}{6}} m a^{2} t^{3} + \chi(t) \right] {\rm e}^{{\rm i} {\bf p} \cdot {\bf a} t^{2}/2 \hbar} |{\bf p}_{0} - m {\bf a} t \rangle  \\
& = & \exp ({\rm i}/\hbar) \left[ {\textstyle \frac{1}{6}} m a^{2} t^{3} + \chi(t) \right] {\rm e}^{{\rm i} {\bf p} \cdot {\bf a} t^{2}/2 \hbar} \, {\rm e}^{- {\rm i} m {\bf a} t \cdot {\bf x}/\hbar} |{\bf p}_{0} \rangle,
\end{eqnarray*}
hence
\[
U^{-1} = \exp ({\rm i}/\hbar) \left[ {\textstyle \frac{1}{6}} m a^{2} t^{3} + \chi(t) \right] {\rm e}^{{\rm i} {\bf p} \cdot {\bf a} t^{2}/2 \hbar} \, {\rm e}^{- {\rm i} m {\bf a} t \cdot {\bf x}/\hbar}.
\]
Thus, from Eq.\ (\ref{3.3}), making use of (\ref{2.2.2}), we obtain
\begin{equation}
K = U H U^{-1} + {\bf a} t \cdot {\bf p} - m {\bf a} \cdot {\bf x} - \frac{1}{2} m a^{2} t^{2} + \frac{{\rm d} \chi}{{\rm d} t}. \label{h2.4}
\end{equation}
If we take $H = {\bf p}^{2}/2m$, corresponding to a free particle, we have
\begin{eqnarray*}
K & = & \frac{({\bf p} - m {\bf a} t)^{2}}{2m} + {\bf a} t \cdot {\bf p} - m {\bf a} \cdot {\bf x} - \frac{1}{2} m a^{2} t^{2} + \frac{{\rm d} \chi}{{\rm d} t} \\
& = & \frac{{\bf p}^{2}}{2m}  - m {\bf a} \cdot {\bf x} + \frac{{\rm d} \chi}{{\rm d} t}.
\end{eqnarray*}
Choosing $\chi = 0$, the Hamiltonian $K$ corresponds to a particle in a uniform force field of intensity $m {\bf a}$ [cf.\ Eq.\ (\ref{unif})], and Eqs.\ (\ref{wftr}) and (\ref{ab2.4}) reproduce the result of Ref.\ \cite{vG}.


\section{Connection with classical mechanics}
In this section we shall show that the function $\alpha$ obtained in the examples of Section 2 coincides with the function $F_{1}$ defined by
\begin{equation}
P_{i} {\rm d} X_{i} - H {\rm d} t - (p_{i} {\rm d} x_{i} - K {\rm d} t) = {\rm d} F_{1}, \label{ct}
\end{equation}
where $H = H(X_{i}, P_{i}, t)$ and $K(x_{i}, p_{i}, t)$ are the Hamiltonian {\em functions}\/ for the canonical coordinates $(X_{i}, P_{i})$ and $(x_{i}, p_{i})$, respectively. As is well known, the transformation that relates the coordinates $(X_{i}, P_{i}, t)$ and $(x_{i}, p_{i}, t)$ of the extended phase space is canonical if and only if there exists a function $F_{1}$ such that Eq.\ (\ref{ct}) holds. (Very often, the function $F_{1}$ is called a generating function of the transformation, but that name is not always adequate, as in all the cases considered here, see, e.g., Refs.\ \cite{CT,HM}.)

In the case considered in Section 2.1, Eq.\ (\ref{ct}) takes the form
\[
{\bf p} \cdot {\rm d} {\bf x} - H {\rm d} t - {\bf p} \cdot {\rm d} {\bf x} + K {\rm d} t = {\rm d} F_{1},
\]
i.e., $(K - H) {\rm d} t = {\rm d} F_{1}$, which is equivalent to saying that $K - H$ is some function of $t$ only; hence, $F_{1}$ is some function, $\chi(t)$ [cf.\ Eq.\ (\ref{ab2.1})], and
\[
K({\bf x}, {\bf p}, t) - H({\bf X}, {\bf P}, t) = \frac{{\rm d} \chi}{{\rm d} t}
\]
[cf.\ Eq.\ (\ref{h2.1})]. If $H$ is given by Eq.\ (\ref{unif}), then
\begin{eqnarray*}
K({\bf x}, {\bf p}, t) & = & \frac{{\bf P}^{2}}{2m} - {\bf F} \cdot {\bf X} + \frac{{\rm d} \chi}{{\rm d} t} \\
& = & \frac{{\bf p}^{2}}{2m} - {\bf F} \cdot ({\bf x} - {\bf a}) + \frac{{\rm d} \chi}{{\rm d} t},
\end{eqnarray*}
which reduces to ${\bf p}^{2}/2m - {\bf F} \cdot {\bf x}$ if $\chi = - {\bf F} \cdot {\bf a} t$ (cf.\ Sect.\ 2.1).

In the case of the translations in the momentum (Sect.\ 2.2), Eq.\ (\ref{ct}) yields
\[
({\bf p} - {\bf b}) \cdot {\rm d} {\bf x} - H {\rm d} t - ({\bf p} \cdot {\rm d} {\bf x} - K {\rm d} t) = {\rm d} F_{1}
\]
or
\[
(K - H) {\rm d} t = {\rm d} (F_{1} + {\bf b} \cdot {\bf x}),
\]
which is equivalent to the existence of a function $\chi(t)$ such that $F_{1} + {\bf b} \cdot {\bf x} = \chi(t)$ and $K - H = {\rm d} \chi/{\rm d} t$. Thus, $F_{1} = - {\bf b} \cdot {\bf x} + \chi(t)$, which coincides with the expression for $\alpha$ given in (\ref{ab2.2}).

For the Galilean transformations, considered in Section 2.3, from Eq.\ (\ref{ct}) we have
\[
({\bf p} - m {\bf V}) \cdot {\rm d} ({\bf x} - {\bf V} t) - H {\rm d} t - ({\bf p} \cdot {\rm d} {\bf x} - K {\rm d} t) = {\rm d} F_{1},
\]
i.e.,
\[
- {\bf p} \cdot {\bf V} {\rm d} t -  m {\bf V} \cdot {\rm d} {\bf x} + m V^{2} {\rm d} t + (K - H) {\rm d} t = {\rm d} F_{1},
\]
which can be written in the form
\[
(K - H - {\bf p} \cdot {\bf V} + {\textstyle \frac{1}{2}} m V^{2}) {\rm d} t = {\rm d} (F_{1} - {\textstyle \frac{1}{2}} m V^{2} t + m {\bf V} \cdot {\bf x}).
\]
Thus, there exists a function $\chi(t)$ such that
\[
F_{1} = {\textstyle \frac{1}{2}} m V^{2} t - m {\bf V} \cdot {\bf x} + \chi(t)
\]
[cf.\ Eq.\ (\ref{ab2.3})] and
\[
K = H + {\bf p} \cdot {\bf V} - \frac{1}{2} m V^{2} + \frac{{\rm d} \chi}{{\rm d} t}
\]
[cf.\ Eq.\ (\ref{h2.3})]. If $H({\bf X}, {\bf P}, t) = {\bf P}^{2}/2m$, then
\begin{eqnarray*}
K({\bf x}, {\bf p}, t) & = & \frac{({\bf p} - m {\bf V})^{2}}{2m} + {\bf p} \cdot {\bf V} - \frac{1}{2} m V^{2} + \frac{{\rm d} \chi}{{\rm d} t} \\
& = & \frac{{\bf p}^{2}}{2m} + \frac{{\rm d} \chi}{{\rm d} t},
\end{eqnarray*}
which coincides with $H({\bf x}, {\bf p}, t)$ if $\chi = 0$.

In the case of uniform acceleration considered in Section 2.4, from Eq.\ (\ref{ct}) we have
\[
({\bf p} - m {\bf a} t) \cdot {\rm d} ({\bf x} - {\textstyle \frac{1}{2}} {\bf a} t^{2}) - H {\rm d} t - ({\bf p} \cdot {\rm d} {\bf x} - K {\rm d} t) = {\rm d} F_{1},
\]
or, equivalently,
\[
(K - H - {\bf p} \cdot {\bf a} t + {\textstyle \frac{1}{2}} m a^{2} t^{2} + m {\bf a} \cdot {\bf x}) {\rm d} t = {\rm d} (F_{1} + m {\bf a} \cdot {\bf x} t - {\textstyle \frac{1}{6}} m a^{2} t^{3}).
\]
Hence, there exists a function $\chi(t)$ such that
\[
F_{1} = - m {\bf a} \cdot {\bf x} t + {\textstyle \frac{1}{6}} m a^{2} t^{3} + \chi(t)
\]
[cf.\ Eq.\ (\ref{ab2.4})] and
\[
K({\bf x}, {\bf p}, t) = H({\bf X}, {\bf P}, t) + {\bf p} \cdot {\bf a} t - \frac{1}{2} m a^{2} t^{2} - m {\bf a} \cdot {\bf x} + \frac{{\rm d} \chi}{{\rm d} t}
\]
[cf.\ Eq.\ (\ref{h2.4})]. Taking $H({\bf X}, {\bf P}, t) = {\bf P}^{2}/2m$, corresponding to a free particle, setting $\chi = 0$ we obtain $K({\bf x}, {\bf p}, t) = {\bf p}^{2}/2m - m {\bf a} \cdot {\bf x}$, corresponding to a particle in a uniform force field.

It may be noticed that, in the derivations presented so far in this section, only the function $\alpha$ appears, without reference to $\beta$. However, we can see that Eq.\ (\ref{bas}) amounts to
\[
\beta = \alpha + {\bf p} \cdot {\bf x} - {\bf P} \cdot {\bf X},
\]
which shows that $\beta$ is a ``type $F_{4}$ generating function'' (though, in the examples considered here, it is not really a generating function owing to the fact that the variables ${\bf p}$ and ${\bf P}$ are not functionally independent).

As shown in Ref.\ \cite{PP} (see also Ref.\ \cite{HM}), under a canonical transformation relating the coordinates $(X_{i}, P_{i}, t)$ and $(x_{i}, p_{i}, t)$, the principal function transforms according to
\begin{equation}
S' = S - F_{1},
\end{equation}
in the sense that if the function $S$ is a solution of the Hamilton--Jacobi (HJ) equation for $H$, then $S' = S - F_{1}$ is a solution of the HJ equation for the Hamiltonian $K$, with $H$ and $K$ related as in (\ref{ct}). Thus, at least in the examples considered here, the transformation law for the wavefunctions is related in a simple manner with the transformation law for the Hamilton principal function. This behavior is not totally surprising if we take into account the relationship between the solutions of the Schr\"odinger equation and $\exp {\rm i} S/\hbar$, where $S$ is a solution of the corresponding HJ equation.

By contrast with the assertion in Ref.\ \cite{vG}, we see that the function $\alpha$ (denoted as $- \hbar S$ in Ref.\ \cite{vG}) is not a Hamilton's principal function, but the difference between two of such functions. In fact, the assertion in Ref.\ \cite{vG} (suggested by one of the referees of that paper) simply makes no sense because there are two Hamiltonians (or Lagrangians) involved, one of them corresponding to a free particle and the other to a particle in a uniform force, while a Hamilton's principal function is associated with just one Hamiltonian (or Lagrangian).

\section{Concluding remark}
The examples presented in this paper explicitly show that the representation on the state vectors of a transformation is not completely specified by its action on the coordinates and momenta. The remaining phase factor in the operator $U$ determines (or is determined by) the difference between the Hamiltonians $H$ and $K$.

\end{document}